\documentstyle[aps,twocolumn,epsfig]{revtex}

\title{BEAM COUPLING IMPEDANCES OF OBSTACLES \\
PROTRUDING INTO BEAM PIPE} 

\author{Sergey~S.~Kurennoy \\
        LANSCE-1, Los Alamos National Laboratory, MS H808,  
        Los Alamos, NM 87545, USA }

\begin{document}
\maketitle

\begin{abstract} 
The beam coupling impedances of small obstacles protruding inside
the vacuum chamber of an accelerator are calculated analytically at 
frequencies for which the wavelength is large compared to a typical 
size of the obstacle. Simple formulas for a few important particular 
cases, including both essentially three-dimensional objects like a 
post or a mask and axisymmetric irises, are presented. The analytical 
results are compared and agree with three-dimensional computer 
simulations.
These results allow simple practical estimates of the broad-band 
impedance contributions from such discontinuities.
\end{abstract} 

\section{Introduction} 

Due to high currents in modern accelerators and colliders, even
contributions from small discontinuities of the vacuum chamber
to the impedance budget of the machine have to be accounted for.
Numerous pumping holes --- a few hundred per meter --- in the
vacuum screen of the Large Hadron Collider (LHC) can
serve as an example. Their total contribution to the machine
impedance in the initial design was calculated \cite{SK,RLG} 
using the Bethe theory of diffraction of EM-waves 
by a small hole in a metal plane \cite{Bethe}, and found
to be dangerously large, close to the beam instability threshold.
Proposed design changes reduced the impedance more than
an order of magnitude \cite{SK95}. The method's basic idea is 
that the hole, at frequencies where the wavelength is large 
compared to the typical hole size, can be replaced by two induced 
dipoles, an electric and a magnetic one. Using essentially the same 
idea, the method was extended for arbitrary-shaped small 
discontinuities on the pipe of an arbitrary-shaped cross 
section \cite{KGS}. 
The impedance calculation for a given small discontinuity was 
therefore reduced to finding its electric and magnetic 
polarizabilities.
Analytical results in this direction have been obtained for 
axisymmetric obstacles \cite{K&S}, as well as for holes and 
slots: circular \cite{Bethe} and elliptic \cite{Collin} holes in a 
zero-thickness wall, circular \cite{G&D} and elliptic \cite{R&G} 
holes in a thick wall, various slots \cite{SK93}, and a ring-shaped 
cut \cite{SK96}.

In a recent paper \cite{SK97} the method was applied to calculate 
the coupling impedances of obstacles protruding inside the beam pipe, 
like a narrow post or a mask intercepting synchrotron radiation. 
Formulas derived make practical estimates very simple. Numerical 
simulations required to obtain similar results are necessarily 
3D ones, and therefore are rather involved. This statement is 
generally applicable for any small discontinuities,
but especially for those protruding into the vacuum chamber. Below 
we list the analytical results \cite{SK97} and compare them
with simulations.

\section{General Solution} 

The longitudinal coupling 
impedance of a small discontinuity on the wall of a circular beam 
pipe of radius $R$ is \cite{SK}
\begin{equation}
Z(k) = - i Z_0 k \frac{\alpha_e + \alpha_m}{4 \pi^2 R^2} \ , \label{Z}
\end{equation}
when the wavelength $2\pi/k$ is large compared to the obstacle size. 
Here $Z_0 = 120 \pi$~$\Omega$ is the impedance of free space, 
$k=\omega/c$ is the wave number, and $\alpha_e, \alpha_m$ 
are the electric and magnetic polarizabilities of the discontinuity. 
The transverse impedance is proportional to the same combination 
of polarizabilities $\alpha_e + \alpha_m$, and the real part of 
the impedance is small at such frequencies (see \cite{KGS,SK95} for 
detail, as well as for other chamber cross sections). 
Let the obstacle shape be a half-ellipsoid with semiaxis $a$ in 
the longitudinal direction (along the chamber axis), 
$b$ in the radial direction, and $c$ in the azimuthal one. 
When $a, b, c \ll R$, the obstacle is small and the Bethe approach 
can be applied. To find the polarizabilities, 
one needs to calculate the induced electric dipole 
moment $P$ of the obstacle illuminated by a homogeneous radial 
electric field $E_0$, and the magnetic dipole moment $M$ when it is 
illuminated by an azimuthal magnetic field $H_0$. This problem
was reduced \cite{SK97} to the well-known problem for an ellipsoid 
immersed in a homogeneous field, e.g., \cite{Collin}. 
Adding obvious symmetry considerations, we get
\begin{equation}
\alpha_e = \frac{P}{2\varepsilon_0 E_0} = 
 \frac{2 \pi a b c}{3 I_b} \ , \qquad 
\alpha_m = \frac{M}{2 H_0} = 
 \frac{2 \pi a b c}{3 (I_c - 1)} \ ,       \label{aem}
\end{equation}
where
\begin{equation}
I_b = \frac{abc}{2} \int_0^\infty \frac{ds}
 { (s+b^2)^{3/2}(s+a^2)^{1/2}(s+c^2)^{1/2} } \ ,  \label{I}
\end{equation}
and $I_c$ is given by Eq.~(\ref{I}) with $b$ and $c$ interchanged.

\section{Post and Mask} 

In the case $a=c$, $b=h$ we have an ellipsoid of revolution, and 
the integral in Eq.~(\ref{I}) can be expressed in terms of the 
hypergeometric function ${}_2F_1$:
\begin{equation}
\alpha_e =  \frac{2 \pi a^2 h}
              {{}_2F_1(1,1;5/2;1-h^2/a^2)} \ ,      \label{ae1}
\end{equation}
and 
\begin{equation}
\alpha_m =  \frac{2 \pi a^2 h}
 {{}_2F_1(1,1/2;5/2;1-a^2/h^2) - 3 } \ . \label{am1}
\end{equation}
In the limit $a \ll h$, which corresponds to a pin-like obstacle,  
$\alpha_e \simeq (2\pi h^3/3)/ \left[\ln\,(2h/a)-1 \right] $
 is much larger than $\alpha_m \simeq - 4\pi a^2 h/3$. 
Note that in this limit 
$\alpha_m \simeq -2V$, where $V=2\pi a^2 h/3$ is the volume occupied
by the obstacle (and subtracted from that occupied by the beam 
magnetic field), somewhat similarly to the axisymmetric case \cite{K&S}. 
These results give us a simple expression for the inductive 
impedance of an {\bf narrow pin (post)} of height $h$ and radius 
$a$, $a \ll h$, protruding radially into the beam pipe:\footnote{One 
could use the known result for the induced electric dipole of a narrow
cylinder parallel to the electric field \cite{Landau}. It will only
change $\ln(2h/a) - 1$ in Eq.~(\ref{Zpin}) to $\ln(4h/a) - 7/3$.}
\begin{equation}
Z(k) \simeq - i k Z_0 \frac{h^3}
     {6 \pi R^2 \left(\ln\,(2h/a) - 1 \right) } \ .    \label{Zpin}
\end{equation}
 
One more particular case of interest here is $h=a$, i.e. a 
{\bf semispherical obstacle} of radius $a$. From 
Eqs.~(\ref{ae1})-(\ref{am1}) the impedance of such a 
discontinuity is
\begin{equation}
Z(k) = - i k Z_0 \frac{a^3} {4 \pi R^2 } \ ,         \label{Zss}
\end{equation}
which is $3\pi/2$ times that for a circular hole of the same radius
in a thin wall \cite{SK}.

The MAFIA code package \cite{MAFIA} was used to compute the impedances 
of various small protrusions for comparison with analytical results.
Calculating wakes due to protrusions is a more difficult task for 
MAFIA than those due to cavities. One has to either use a long pipe 
(which leads to a huge mesh since it should be homogeneous 
along the beam path), or apply a trick with tapers on the pipe ends.
The tapers make transitions to a new end pipe with radius $R_p 
 \le R-h$, so that the new structure looks like a shallow cavity. 
The difference of wakes computed with and without a protrusion
gives us its contribution. We used the second approach with 
pipes from $4R$ to $8R$ long and meshes up to $2\cdot10^6$ points.
Simulation results are usually higher than the analytical ones, 
but go down as a finer mesh is used. Figure~1 gives 
some comparison for a semisphere. 

\begin{figure}[htb]
\centerline{\epsfig{figure=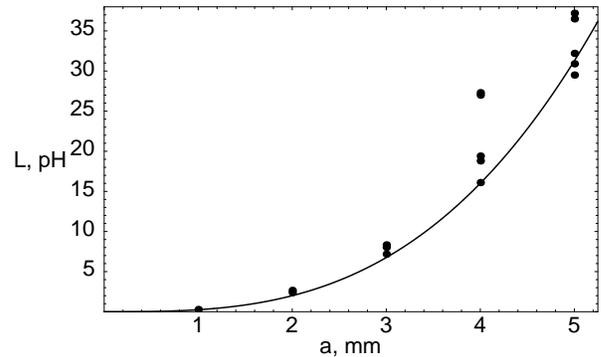,width=8cm}}
\caption{Comparison of MAFIA results (dots) for the inductance of 
a semispherical protrusion with the analytical ones (solid line) 
from Eq.~(\ref{Zss}), for different sphere radii. Pipe radius 
$R=2$~cm.}
\end{figure}

Another practical result that can be derived from the general solution,
Eqs.~(\ref{aem})-(\ref{I}), is the impedance of a {\bf mask} intended to
intercept synchrotron radiation. We put $b=c=h$, so that our model mask
has the semicircular shape with radius $h$ in its largest transverse 
cross section. Then the integral in Eq.~(\ref{I}) is reduced to 
$$I_b=I_c= {}_2F_1 \left(1,1/2;5/2;
 1-h^2/a^2\right)/3 \ ,$$ 
and we further simplify the result for two particular cases. 

The first one is the {\bf thin mask}, $a \ll h$,
in which case $\alpha_e \simeq  8h^3/3$ , and again it dominates the
magnetic term, $\alpha_m \simeq  -V = -2\pi a h^2/3$. The coupling 
impedance for such an obstacle 
--- a half-disk of radius $h$ and thickness $2a$, $a \ll h$, 
transverse to the chamber axis --- is therefore
\begin{equation}
Z(k) = - i k Z_0 \frac{2 h^3} {3 \pi^2 R^2 } \left [ 1 + 
  \left( \frac{4}{\pi} - \frac{\pi}{4} \right) \frac{a}{h}
  + \ldots \, \right ] \, ,        \label{Zm0}
\end{equation}
where the next-to-leading term is shown explicitly. 

In the opposite limit, $h \ll a$, which corresponds to a {\bf long} 
(along the beam) {\bf mask}, the leading terms 
$\alpha_e \simeq - \alpha_m \simeq  4\pi a h^2/3$ 
cancel each other. As a result, the impedance 
of a long mask with length $l=2a$ and height $h$,  $h \ll l$, is
\begin{equation}
Z(k) \simeq - i k Z_0 \frac{4 h^4}{3 \pi R^2 l } 
    \left(\ln \frac{l}{h} - 1 \right) \ ,               \label{Zml}
\end{equation}
which is relatively small due to the ``aerodynamic'' shape of this 
obstacle, in complete analogy with results for long elliptic slots 
\cite{SK,RLG,SK93}. 
Figure~2 shows the impedance of a mask with a semicircular 
transverse cross section of radius $h$ versus its normalized 
half-length, $a/h$. The comparison with the asymptotic approximations 
Eqs.~(\ref{Zm0}) and (\ref{Zml}) is also shown. One can see that
the asymptotic behavior (\ref{Zml}) starts to work well only for very
long masks, namely, when $l = 2a \ge 10h$. Figure~2 demonstrates
that the mask impedance depends rather weakly on the length. 
Even a very thin mask ($a \ll h$) contributes as much as $8/(3\pi) 
 \simeq 0.85$ times the semisphere ($a=h$) impedance, Eq.~(\ref{Zss}),
while for long masks the impedance decreases slowly: at $l/h=20$, it
is still 0.54 of that for the semisphere. 

\begin{figure}[htb]
\centerline{\epsfig{figure=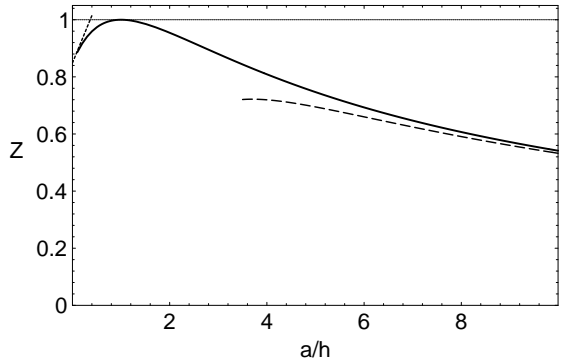,width=8cm}}
\caption{Mask impedance $Z$ (in units of that for a semisphere 
with the same depth, Eq.~(\ref{Zss}) with $a=h$) versus its length. 
The narrow-mask approximation, Eq.~(\ref{Zm0}), is short-dashed, 
and the long-mask one, Eq.~(\ref{Zml}), is long-dashed.}
\end{figure}
 
In practice, however, the mask has usually an abrupt cut toward the 
incident synchrotron radiation, so that it is rather one-half of a
long mask. Numerical simulations show that the low-frequency 
impedances of a semisphere and a half-semisphere of the same depth 
 --- which is a relatively short mask --- are almost equal 
(within the errors), and close to that for a 
longer half-mask. From these results one can conclude 
that a good estimate for the mask impedance is given simply
by Eq.~(\ref{Zss}). One more interesting observation is that 
the inductances found numerically for two half-semispheres --- 
one with a cut toward the incident beam and the other, mirrored 
$z \to -z$ --- differ by 3\%, while they have to be equal 
as was proven analytically \cite{G&Z&SH}. 

\section{Axisymmetric Iris} 

Following a similar procedure one can also easily obtain the 
results for axisymmetric irises having a semi-elliptic profile 
in the longitudinal chamber cross section, 
with depth $b=h$ and length $2a$ along the beam. 
For that purpose, one should consider the limit $c \to \infty$ in
Eq.~(\ref{I}) to calculate polarizabilities $\tilde{\alpha}_e$ and 
$\tilde{\alpha}_m$ per unit length of the circumference of the
chamber transverse cross section. The broad-band impedances
of axisymmetric discontinuities have been studied in \cite{K&S},
and the longitudinal coupling impedance is given by
\begin{equation}
Z(k) = - i Z_0 k \frac{\tilde{\alpha}_e + 
         \tilde{\alpha}_m}{2 \pi R} \ ,                \label{Zax}
\end{equation}
quite similar to Eq.~(\ref{Z}). As $c \to \infty$, the integral
$I_c \to 0$, and $I_b$ is expressed in elementary functions as
$$I_b=\frac{1}{2}{}_2F_1 \left(1,\frac{1}{2};2;
 1-\frac{h^2}{a^2}\right) =\frac{a}{a+h} \ . $$
It gives us immediately
\begin{equation}
 \tilde{\alpha}_e = \pi h(h+a)/2 \, ; \qquad
  \tilde{\alpha}_m = - \pi ah/2 \ ,         \label{alax}
\end{equation}
and the resulting impedance of the iris of depth $h$ with the 
semielliptic profile is simply
\begin{equation}
Z(k) = - i k Z_0 \frac{h^2} {4R } \ ,               \label{Ziris}
\end{equation}
which proves to be independent of the iris thickness $a$. The same
result has been recently obtained using another, direct 
method \cite{RLGu}. 

One should emphasize that $\tilde{\alpha}_m$ in 
Eq.~(\ref{alax}) is just an iris cross-section area (with negative
sign), which is correct for any small axisymmetric discontinuity, 
as was pointed out in \cite{K&S}. However, calculating 
$\tilde{\alpha}_e$ in general is not easy: a conformal mapping was 
constructed for that purpose in \cite{K&S} for irises (as well as 
for chamber enlargements) having a trapezoid (or rectangular, or 
triangular) profile. An interesting fact is that the leading behavior 
for thin irises of all shapes is exactly the same as Eq.~(\ref{Ziris}). 

In fact, the same approach can be applied here. The conformal mapping 
from the upper half-plane $w$ into $z$ with the boundary including 
the iris having a semielliptic profile is given by 
$$ 
z = aw + h \sqrt{\, w^2 -1} \ . 
$$
We need an inverse mapping, but, fortunately, it is enough to find 
its asymptotic behavior at large $z$ and $w$ \cite{K&S}, which is 
$$ 
w = \frac{z}{a+h} + \frac{h}{2}\, \frac{1}{z} + \ldots 
$$
The ratio of the coefficients of the second and 
first terms on the RHS is $\tilde{\alpha}_e/\pi$, 
cf.\ \cite{K&S}, which leads us exactly to 
the result for $\tilde{\alpha}_e$ in Eq.~(\ref{alax}).
It is even easier for the particular case $h=a$, when the iris has 
a semicircular profile of radius $a$. The explicit conformal mapping 
for this case is very simple, $w=(z/a+a/z)/2$. The comparison of the 
second and first terms on the RHS gives us 
$\tilde{\alpha}_e  = \pi a^2$, in agreement with Eq.~(\ref{alax}).

\section{Summary}

Analytical expressions of the impedances for both 3D and axisymmetric 
small obstacles protruding inside a beam pipe are obtained, and
they agree well with numerical results. These formulas greatly 
simplify calculations of the broad-band contributions to the 
coupling impedances from such discontinuities, especially in the
3D case. 

 The present approach does not work for enlargements 
of the vacuum chamber. However, existing analytical results for holes 
and slots \cite{SK,RLG,SK95,KGS}, as well as for axisymmetric 
enlargements \cite{K&S}, cover this case quite well.


Stimulating discussions with Dr.\ R.L.~Gluckstern and 
Dr.\ F.L.~Krawczyk are gratefully acknowledged.

\end{document}